\documentstyle[maps,epsf, twocolumn,aps,floats]{revtex}  

\begin{document}
\draft
\author{B. Utter, R. Ragnarsson and  E. Bodenschatz}
\address{Laboratory of Atomic and Solid State Physics, Cornell University,
    Ithaca, NY  14853}
\date{\today}
\title{Quasiperiodic Tip Splitting in Directional Solidification}

\mabstract{ We report experimental results on the tip splitting
dynamics of seaweed growth in directional solidification of
succinonitrile alloys with poly(ethylene oxide) or  acetone as
solutes. The seaweed or dense branching  morphology was selected by
solidifying grains which are oriented close to the $\{$111$\}$ plane.
Despite the random appearance of the growth, a quasiperiodic tip
splitting morphology was observed in which the tip alternately splits
to the left and to the right.  The tip splitting frequency $f$ was
found to be related to the growth velocity $V$ as a power law $f
\propto V^{1.5}$. This finding is consistent with the predictions of a
tip splitting model that is also presented. Small anisotropies are shown to 
lead to different kinds of seaweed morphologies. }


\mmaketitle             



Directional solidification of binary alloys has received considerable
attention over the last fifty years as an example of a non-equilibrium pattern
forming system with technological importance. A breakthrough in understanding
was gained when surface tension anisotropy was identified as a singular
perturbation that selects stable dendrites 
\cite{Kessler.ea:85:Geometrical,Ben-Jacob.ea:84:Pattern}. 
Many important
experiments have been performed to understand the structure and dynamics of
dendritic growth and great insights have been gained 
\cite{Kurz.ea:90:Solidification}. 
However, when
growing a cubic crystal close to the $\{111\}$ plane, the surface
tension is nearly isotropic and it was shown
\cite{ihle-mk-93-4,Akamatsu-95-fig20} that the growth is
irregular with constant tip splitting. This is called seaweed growth
\cite{ihle-mk-93-4}, or dense branching morphology
\cite{Ben-Jacob.ea:86:Formation}, and is also observed in such different
systems as viscous fingering \cite{McCloud.ea:95:Experimental}, bacterial
colonies \cite{Matsuyama.ea:93:Fractal}, electrodeposition
\cite{Zik.ea:96:Electrodeposition}, annealing magnetic films
\cite{Shang:96:Pattern-formation}, and drying water films
\cite{Samid-Merzel.ea:98:Pattern}.

In this Letter, we report experimental results on the dynamics of seaweed
growth in the directional solidification of binary succinonitrile alloys. By
using a grain oriented near the $\{111\}$ plane, we observe quasiperiodic
alternating tip splitting \cite{Akamatsu-95-fig20}. We find the tip
splitting frequency to be related to the growth velocity by a power law 
with an exponent close to 1.5. We show
that a simple model motivated by observations of the growth dynamics can
explain this behavior and that a small misalignment of the
crystal from the $\{111\}$ plane is necessary for quasiperiodic 
alternating tip splitting.
To our knowledge, neither the scaling nor the regularly alternating splitting
have been predicted theoretically. Our results show that different types of
seaweed morphologies exist that at first sight would appear the same.

\begin{figure}[bt]
\epsfxsize=3.4in
\begin{center}\mbox{
\epsffile{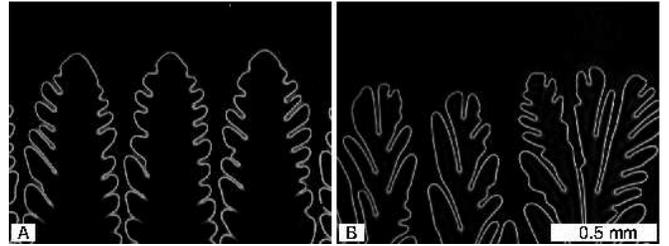} }\end{center} \caption{ Phase
contrast micrographs of growth in directional solidification which differ only
in crystalline orientation. (A) A dendrite (growth along $\{100\}$ direction) 
and (B)
seaweed (near $\{111\}$ plane). The white line indicates the solid-liquid
interface with the solid growing upwards into the undercooled melt.  The 
thermal
gradient (18 K/cm), concentration (0.25$\%$ PEO), and growth velocity V (2.71
$\mu$m/s) are the same.} \label{seaweed}
\end{figure}

The surface tension anisotropy is based on the underlying cubic structure of
the growing solid.  Under typical conditions in low speed directional
solidification, the orientation of the dendrites and sidebranches is
determined by the crystalline structure, pulling direction,  and
imposed temperature gradient. Changing the angle between the crystalline axis
and the pulling direction changes the degree and orientation of the effective
anisotropy \cite{Akamatsu-95-fig20,Akamatsu.ea:98:Anisotropy-driven}. 
Because of the cubic
symmetry, growth in the $\{111\}$ plane has an effective surface tension that
is close to isotropic. As shown in Fig.~\ref{seaweed}, under the same growth
conditions, dendrites are observed for crystals growing in the $\{100\}$
direction and disordered seaweeds with continuous tip splitting are observed
for crystals oriented in the $\{111\}$ plane.

The present experiment was performed with a directional solidification
apparatus in which a thin sample $(13$cm x $1.5$cm x (5-50)$\mu$m) was moved
through a linear temperature gradient at constant velocity
\cite{Ragnarsson:1998:Super}. 
In these experiments, the gradient was maintained at about
18 K/cm with a stability of $\pm 10$ mK on each side. The two mixtures 
used were transparent alloys of 
0.25$\%$ poly(ethylene oxide) (PEO) \cite{peo-info} and 1.5$\%$ acetone (ACE) 
in succinonitrile (SCN). The polymer has a very small partition coefficient and
diffusivity ($k \approx 0.01, D \approx 80 \mu$m$^2/$s) while for the acetone
sample they are larger $(k \approx 0.33, D \approx 1300 \mu$m$^2/$s)
\cite{Chopra.ea:88:Dendritic}.

The solid-liquid interface is observed with phase contrast microscopy, and
images are recorded using a CCD camera and time lapse video. To
observe seaweed growth, the sample is melted and quenched, seeding
numerous grains. One grain with the optimal orientation is selected and all
the others are melted off.  The chosen grain is then allowed to grow and fill 
the width of the cell.

For seaweed growth, we observe two types of tip splitting events:
(i) The tip splits off-center such that the larger of the two remaining tips
continues to grow as the other falls behind. 
This can lead to growth in which
the tip alternately splits towards the left and right at regular intervals as
a single main branch steadily grows forward. (ii) The tip splits near its
center producing two tips of comparable size.  These two new branches 
interact with each other until either one falls behind, or they spread
apart and grow as main branches. 
In this Letter, we will focus on the dominant case (i) and specifically
on quasiperiodic
alternating tip splitting.

\begin{figure}[tb]
\epsfxsize=3.0 in
\begin{center}\mbox{
\epsffile{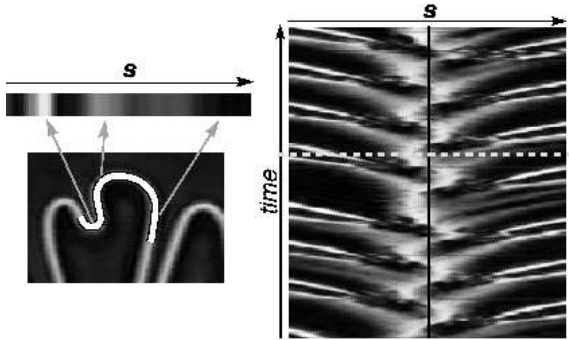} }\end{center} 
\caption{
Curvature-time (CT) plots for SCN-PEO. 
A represention of the curvatures along the
interface near the tip. 
The solid-liquid interface is extracted and 
the absolute curvatures along a segment near the tip are plotted in
greyscale. 
Sequences of these lines are stacked in time
to create the plot on the right, where the center line represents the 
tip position. The 
tip segment highlighted 
on the left is indicated as a dashed line on the CT plot. 
White corresponds to high curvature (radius of curvature $<$ 10
$\mu$m) and black to zero curvature.
The time interval shown is 9.4 minutes and increases upwards. The width is 
300 $\mu$m and $V = 2.71 \mu$m/s. } \label{crvprf}
\end{figure}

We characterized the tip splitting process by measuring the curvature at each
point along an arc centered on the tip, 
which we defined as the furthest point along the
growth direction on a particular branch. The segment used is typically 300
$\mu$m long as compared to the tip radii of (25-75)$\mu$m. Plotting the
curvature versus the position along the arc and stacking the plots from
successive times, we created curvature-time (CT) plots 
as shown in Fig.~\ref{crvprf}.  There, the
greyscale intensity corresponds to the absolute value of the curvature.  
The striking feature is that the curvatures at
the tip oscillate.  The CT plot shows how the tip flattens and splits 
on one side before it begins to flatten on the other side.  
The bright pairs of lines to the sides
represent the large curvature at the deep groove formed when a tip splits.
The tip splitting is not random
as one might expect for noise-induced splitting; for most of the runs,
the splitting alternates regularly for more than 85$\%$ of the time. 
It should be noted,
however, that this behavior is quasiperiodic. The time between individual tip
splittings fluctuates within 25$\%$ of the mean period and the tip
occasionally splits multiple times on the same side.

The periodicity is also broken occasionally (typically after 10-20 cycles)
when the tip splits nearly in the center, 
forming two symmetric main branches (case ii). These two
``new'' branches influence each other by initially suppressing the tip
splitting on their adjacent sides, reminiscent of doublon growth
\cite{Brener.ea:94:Fluctuation}. Once one branch falls behind or the tips are
separated sufficiently, 
the tips resume the quasiperiodic alternating tip
splitting.

\begin{figure}[tb]
\epsfxsize=3.0 in
\begin{center}\mbox{
\epsffile{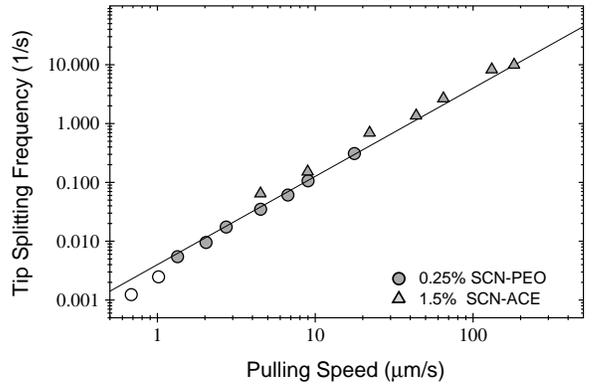} }\end{center} \caption{Tip splitting
frequency $f$ versus growth velocity $V$ for two samples.  The solid line 
shows $f = 0.004 \times
V^{1.5}$.  The best fits for the individual runs give exponents of 1.56
$\pm$ 0.05
for SCN-PEO and 1.40 $\pm$ 0.05 for SCN-ACE.  The two points 
below 
1.1 $\mu$m/s were not used in the fit as they are below a transition to 
cellular growth as discussed later.}
\label{freqvsspeed}
\end{figure}

The splitting frequency was measured at each velocity by following a single 
branch over
long times.  It was determined from the peak of the power spectrum of the
CT plot, 
from the peak of the probability distribution of time intervals between
splitting events, or simply by counting the number of times the tip splits and
dividing by the total time of the run, 
all of which gave the same results within
measurement errors. 
As shown in Fig.~\ref{freqvsspeed}, the splitting frequency as a function of
velocity follows the power law $f \propto V^{1.5}$ for the two mixtures.

\begin{figure}[tb]
\epsfxsize=3.0in
\begin{center}\mbox{
\epsffile{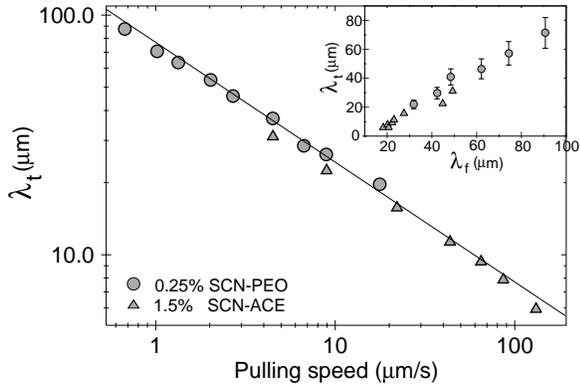} }\end{center} \caption{ 
The wavelength of the instability at the tip $\lambda_{t}$
is plotted versus pulling speed.
The line corresponds to $\lambda_{t} = 75 \times V^{-0.5}$. 
The measured power law
exponents are $-0.46 \pm 0.05$ for SCN-PEO and $-0.48 \pm 0.05$ 
for SCN-ACE. The inset
shows the approximately linear relationship between $\lambda_{t}$ and
$\lambda_{f}$.  For SCN-PEO, $\lambda_{t}(V) = 0.85 \lambda_{f}(V)- 7.4 \mu$m 
and for SCN-ACE, $\lambda_{t}(V) = 0.73 \lambda_{f}(V) - 6.9 \mu$m. } 
\label{SLvsV}
\end{figure}

To model this behavior, we note that the tip 
flattens and widens before becoming unstable. One can
expect that the wavelength  $\lambda_{t}$ of this tip instability is related
to the initial instability of the flat interface $\lambda_{f}$ under the same
conditions. This is confirmed for both samples in 
Fig.~\ref{SLvsV} (inset). 
$\lambda_t$ is smaller than $\lambda_f$ since the expanding tip will
split at the smallest wavelength that is unstable in this 
evolving state.
Fig.~\ref{SLvsV} shows the wavelength $\lambda_{t}$ as a function
of pulling speed for both samples and the solid line corresponds to
$\lambda \propto V^{-0.5}$.
The exponent of $-\frac{1}{2}$ is 
the expected value for $\lambda_{f}$ based on 
linear stability analysis \cite{Mullins.ea:64:Stability}.
In practice, $\lambda_{t}$ is more convenient to measure
since it can be determined as an average over many events during the portion
of the run being studied.  
From dimensional analysis, we now see that the frequency results from 
the pulling velocity $V$ divided by the length scale $\lambda$. 
That is, 
$f \propto V/\lambda \propto V^{-3/2}$ appears to be satisfied.

\begin{figure}[tb]
\epsfxsize=2.8 in
\begin{center}\mbox{
\epsffile{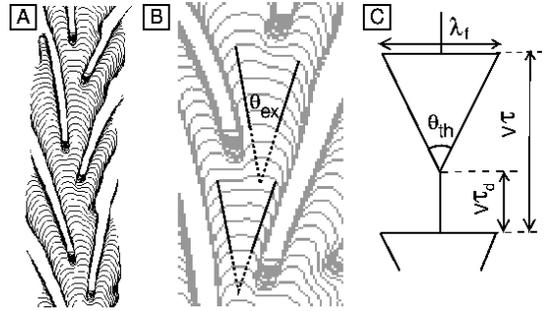} }\end{center} \caption{ (A) The
interface is extracted at sequential times and superimposed to observe the
growing tip.  (B) An approximately flat region of the tip is seen to 
grow linearly in time.  
The boundaries of this expanding flat region (solid lines) 
are traced back (dashed) to the 
same point.  
(C) Schematic of the growth process (see text). 
}
\label{triangle-model}
\end{figure}

Additional quantitative 
insight is gained by extracting the interface at subsequent
times and superimposing the images to observe the growing solid 
(Fig.~\ref{triangle-model}A). It is apparent that the flat
region of the tip grows until it is wide enough to become unstable.
We observe that the tip widens at a
constant rate until it reaches the instability length of the initial flat
interface.  After 
splitting, the surviving tip widens in a similar way. 
As seen in Fig.~\ref{triangle-model}B, we can view this
process as a series of triangles formed by the
flat regions. For SCN-PEO, we often 
observe a delay between consecutive splittings,
which introduces a gap between the triangles in Fig.~\ref{triangle-model}B. 
This may be explained by the fact that 
the tip is initially constrained by the presence
of the neighboring tip and is inhibited from widening until the neighbor
falls behind. 
The average period between splitting is $\tau = 1/f$ 
and we assume the delay time is a fixed fraction of the average period,
$\tau_d  = \alpha \tau$, where $\alpha$ is
constant for a given sample.  If $\alpha$ is small, 
the angle $\theta$ at which the tip widens is predicted from
Fig.~\ref{triangle-model}C to be
\begin{equation}
\theta_{th} = 2 \arctan \left( \frac{\lambda_f f}{2 V} 
\left( 1 + \alpha \right) \right).
\label{theta-eqn}
\end{equation}

\begin{figure}[bt]
\epsfxsize=2.6in
\begin{center}\mbox{
\epsffile{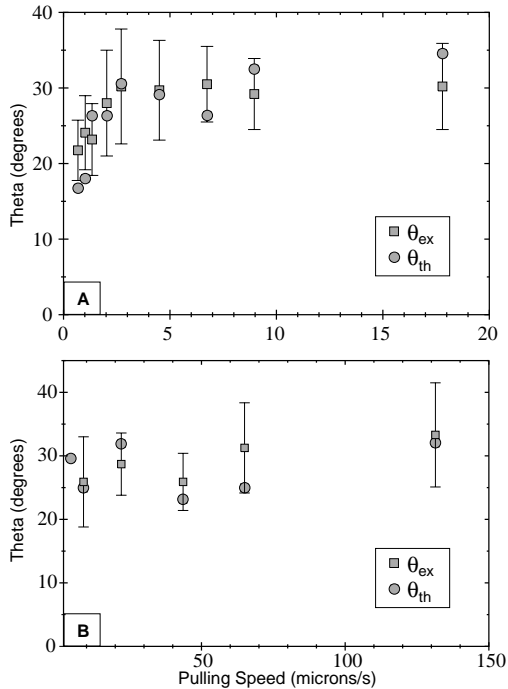} }\end{center} \caption{ 
Comparison of the measured and predicted angles $\theta_{ex}$ and 
$\theta_{th}$.  (A) SCN-PEO with $\alpha$ = 0.37 and (B) SCN-ACE with
$\alpha$ = 0. }
\label{ThetaVSv}
\end{figure}

All quantities on the right, other than the free parameter $\alpha$, 
have been measured, so a quantitative comparison can be made
between the predicted value $\theta_{th}$ and $\theta_{ex}$
which was measured directly from the
experimental pictures.
Fig.~\ref{ThetaVSv} presents this data for the two mixtures considered.  
We observed excellent agreement between the model and the measurement.
In addition, we observe the angle to be 
independent of velocity, which with Eq.~\ref{theta-eqn} 
implies the above scaling relationship, 
$f \propto V/\lambda$.  It also suggests 
that $\theta$ depends only on a materials
property, such as the surface tension of the growing solid. The fact that the
two samples show similar angles may indicate that a similar surface tension
profile exists in these particular grains. The drop off at low speeds for 
SCN-PEO
reflects a change in morphology since when the speed was decreased the
growth became increasingly 
cellular and the tip did not continually split. 
It is unclear why the SCN-ACE sample does not exhibit the 
delay seen in the SCN-PEO mixture.

The above observations suggest that the 
alternating tip splitting may be correlated with specific
crystalline orientations near $\{111\}$ and not with the choice of solute or
sample dimensions.  
In fact, Akamatsu et al. also noted a region of quasiperiodicity 
in a CBr$_4$-C$_2$Cl$_6$ mixture 
which is similar to our observations
\cite{Akamatsu-95-fig20}.
It is important to note that 
it is possible to seed other kinds of seaweeds in these samples which do
not exhibit this striking periodicity. This confirms that the specific grain is
important.  

\begin{figure}[tb]
\epsfxsize=2.5in
\begin{center}\mbox{
\epsffile{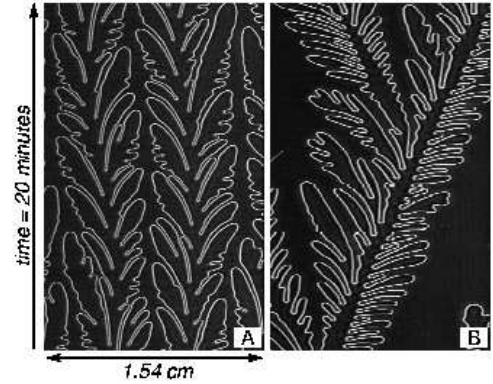} }\end{center} \caption{ 
Space-time (ST) plot
of a single grain growing at 2.71 $\mu$m/s. (A) The quasiperiodic splitting is
observed, but when the temperature gradient is decreased from $18 K/$cm to (B)
$3 K/$cm, the degeneracy of growth is revealed as the growing solid locks into
two symmetric directions.   The total time shown is 20 minutes in
both pictures. } \label{degenerate-ST}
\end{figure}

To support this conjecture,
the temperature gradient was decreased in order to diminish the effect
of the imposed growth direction.
Fig.~\ref{degenerate-ST} compares a space-time plot (ST)
\cite{stplot} of a quasiperiodic 
tip splitting growth (Fig.~\ref{degenerate-ST}A) 
with that for the same grain when the temperature gradient is decreased 
(Fig.~\ref{degenerate-ST}B).  The growth locks into two symmetric
directions revealing the weak anisotropy.   This accounts for
the tendency of the tip to grow outward and flatten as growth along these
directions is preferred.
After converting time into distance, 
the angle between the two growth directions in Fig.~\ref{degenerate-ST}B 
is $40^\circ$, which should be
an upper limit on the angle measured in Fig.~\ref{ThetaVSv}. 
We conclude that a small amount of anisotropy has important consequences
on the seaweed growth observed.  It also confirms that seaweed growth is
possible for weak anisotropy.

In conclusion, experimental studies of the low anisotropy seaweed growth
morphology for certain grains show surprisingly regular tip splitting despite
its generally disordered appearance. We find that the curvature of the tip
oscillates, reflecting the cyclic changes in its shape which lead to
alternating tip splitting. The tip splits quasiperiodically, with a frequency
that is related to the growth velocity as $f \propto V^{3/2}$.  
We present a simple
model that assumes that the tip splits when the width of 
the flat region on the tip
is sufficiently wide for the linear instability to occur.  
The angle at which
the tip widens is independent of growth velocity explaining the observed
scaling relation.
We show evidence that a slight
anisotropy in the surface tension causes the observed alternating 
tip splitting.
It is interesting to ask whether the behavior observed here could also be 
found in other systems, such as Saffman-Taylor fingering where a weak 
anisotropy is imposed. 

We thank Nigel Goldenfeld for valuable discussions.
This work was supported by the Cornell Center for Materials Research (CCMR), a
Materials Research Science and Engineering Center of the National Science
Foundation (DMR-0079992).



%
%

%
%

\end{document}